\documentclass[a4paper,12pt]{article}
\usepackage{latexsym}
\usepackage[T1]{fontenc}
\begin{document}
\title{On the issue of gravitons}
\author{Leszek M. SOKO\L{}OWSKI\thanks{Corresponding author. 
Email: UFLSOKOL@th.if.uj.edu.pl} \\ Astronomical Observatory,
Jagellonian University, Orla
171, \\ 30-244 Krak\'ow, Poland \\ and \\
Andrzej STARUSZKIEWICZ \\ Institute of Physics,
Jagellonian University, \\ Reymonta 4,
30-059 Krak\'ow, Poland}
\date{}
\maketitle
Short title: On the issue of gravitons
\begin{abstract}
We investigate the problem of whether one can anticipate any features of 
the graviton without a detailed knowledge of a full quantum theory of 
gravity. Assuming that in linearized gravity the graviton is in a sense 
similar to the photon, we derive a curious 
large number coincidence between the number of gravitons emitted by a 
solar planet during its orbital period and the number of its constituent 
nucleons (the coincidence is less exact for extra solar planets since 
their sample is observationally biased). The coincidence raises a 
conceptual problem of quantum mechanism of graviton emission and we show 
that the problem has no intuitive solution and there is no physical picture 
of quantum emission from a macroscopic body. In Einstein's general 
relativity the analogy between the graviton and the photon turns out ill 
founded. A generic relationship between 
quanta of a quantum field and plane waves of the corresponding classical 
field is broken in the case of full GR. 
The graviton cannot be classically approximated by a 
generic \emph{pp\/} wave nor by its special case, the exact plane wave. 
Furthermore and most important, the ADM energy is a 
zero frequency characteristic of any asymptotically flat gravitational field, 
this means that any general relationship between energy and frequency is 
\emph{a priori\/} impossible. In particular the formula 
$E=\hbar \omega$ does not hold. The 
graviton must have features different from those of the photon and these 
cannot be predicted from classical general relativity. 
\end{abstract}

\section{Introduction}
The notion of graviton is popular in modern physics even though any version 
of quantum gravity (e.g. loop quantum gravity) is still far from providing a 
well grounded derivation of the concept. The notion is based on pure 
analogy with quantum theory of other fields, notably electromagnetic field. 
According to general rules of QFT quantum fields are the basic ingredients 
of any matter and particles are the quanta, i.e. \/ grains (or bundles) of 
the four--momentum of the fields. The graviton, like the photon, is a fully 
quantum object. In the case of electromagnetic field the photon may be 
approximated, in the low energy limit, by a classical plane monochromatic 
wave and the photon's energy $E$ is related to the wave frequency $\omega$ 
by the Planck--Einstein formula $E=\hbar \omega$ or relativistically, 
$p^{\mu}=\hbar k^{\mu}$, $k^{\mu}$ being the wave four--vector. Historically, 
Einstein and others followed in the opposite direction and associated a 
quantum of the electromagnetic field with a monochromatic plane wave. 
Their conjecture was then fully confirmed in the framework of QED and later 
in other quantum field theories. The conjecture, though being a basis for 
the transition to the quantum theory, is compatible with classical 
electrodynamics and in conjunction with the very concept of a quantum may 
be anticipated in the classical theory. Namely, a general radiation field 
may be decomposed into plane waves and then the total four--momentum 
is an integral over the momentum space of a product of a relativistic 
scalar and $k^{\mu}$; the scalar may be interpreted as the number of 
photons, each with $p^{\mu}=\hbar k^{\mu}$. In this sense the wave--particle 
duality is already grounded in the classical field theory. \\

The concept of graviton has been based on this analogy and it works in 
quantum linearized gravity, though not without reservation, see sect. 4. 
The problem is whether it may also work in full nonlinear general relativity, 
the latter being not a field theory in Minkowski spacetime but a theory of 
the spacetime geometry itself. The well known difficulties with constructing 
a quantum theory of gravity clearly indicate that great caution is needed 
whenever one assigns \emph{a priori\/} any property to the quantum 
gravitational field. It is rather expected that the graviton is drastically 
different from the photon and the quanta of other matter fields. \\

In section 2 we assume that the concept of graviton is legitimate in quantum 
linearized gravity and we apply it to the classical quadrupole formula in 
the case of gravitationally radiating planets orbiting around the Sun. We find 
then a large number coincidence (numbers of order $10^{50}$ for the Earth and 
$10^{54}$ for Jupiter) between the number of radiated away gravitons and the 
number of nucleons in the emitting planet. In the case of eight extrasolar 
planets with known all the orbital parameters necessary to compute the 
radiated power, the large number coincidence is less conspicuous (the two 
numbers may differ by a factor $10^4$) and the divergence is likely to be 
due to an 
observational bias. The coincidence seems to be rather accidental (at 
present there are no hints that it may arise from the first principles), 
nevertheless it is interesting in its own and moreover it raises a problem 
of a physical mechanism responsible for the emission of gravitons by a 
macroscopic body. We discuss the problem of the mechanism in section 3 and 
argue, by analogy with QED, that no physical picture of the quantum 
emission of gravitons (by the individual nucleons? by the whole planet?) 
may be found and one's belief in the correctness of this and other predictions 
of quantized linear gravity is founded solely on making calculations in the 
framework of this theory. \\
Section 4 is the heart of the work. We show there how the concept of photon 
is anticipated in CED and how the analogous reasoning in linearized general 
relativity is plagued with troubles with the notion of the gravitational 
energy density. In full GR the situation is much worse: if the hypothetical 
graviton carries energy $E$ and momentum $\mathbf{p}$ then it cannot be 
classically approximated by the \emph{pp\/} waves since the only value 
of the ADM energy and momentum that may be associated with the waves is 
zero. Furthermore, the ADM energy of any asymptotically flat spacetime  
is effectively a charge (computed at the spatial 
infinity) and thus it cannot be classically related to a frequency 
different from zero. The fundamental relation between energy and wave 
frequency in quantum theory of any matter, $E=\hbar \omega$, is broken in 
quantum gravity. If the gravitational field has a quantum nature (what 
is not so obvious in the light of recent ideas of emergent gravity) and 
its quantum does exist, no properties of the graviton can be 
anticipated from classical general relativity (applying standard quantum 
mechanics) and the analogy with the photon is false. In other terms, if 
quantum gravity effects do exist, the correspondence between quantum and 
classical gravity is more sophisticated than in other quantum fields and 
any signs of the quantum effects appear in Einstein's general relativity 
at places where we cannot imagine them at present. 

\section{The quadrupole formula, gravitons and a large number coincidence}
In the linearized version of general relativity (the background 
spacetime is Minkowski space) the gravitational radiation emitted 
by any isolated system of masses is dominated by the quadrupole 
component. The power of this radiation is given by the
quadrupole formula.
We consider radiation from planets on circular orbits of radius $r$ 
around the Sun. Let $m$ be a planet's mass and $M$ the 
Solar mass. Then the emitted power $P$ of gravitational radiation is 
expressed in terms of the reduced mass $\mu$ of the system and the angular 
velocity of the rotating quadrupole $\omega$, 
\begin{equation}
P = \frac{32G}{5c^5}\mu^2\omega^6 r^4.
\end{equation}
The angular velocity of any planet 
is determined by the third Kepler's law, 
\begin{displaymath}
T = 2\pi\sqrt{\frac{r^3}{G(M+m)}}, 
\end{displaymath}
where $T$ is the planet's period of revolution around the Sun. 
Then eq. (1) takes the form 
\begin{equation}
P = \frac{32G^4}{5c^5}\frac{M^2(M+m)m^2}{r^5}.
\end{equation}
The angular velocity $\omega$ of the rotating quadrupole is equal to the 
frequency of the emitted monochromatic gravitational waves. \\

If the gravitational interaction is fundamentally of quantum nature, 
then according to our interpretation of quantum field theory we expect 
that a classical gravitational wave is actually a bundle of gravitons. 
Here one makes two crucial assumptions. First, that any quantum theory 
of gravity should reduce, in the weak field approximation and under some 
other assumptions (which are presently unknown), to linearized general 
relativity, i.e. the quadrupole formula should be valid in an 
appropriate limiting case. Second, at least in some approximation, the 
general picture of a quantized field as a collection of particles being 
bundles of energy and momentum, applies to quantum gravity and gravitons 
are quanta of the gravitational field. If the two assumptions are valid 
one may view the radiation emitted by each planet as a flux of 
gravitons with frequency $\omega$ and wavelength $\lambda = cT$.  
Each graviton carries energy $\hbar \omega$. 
For the Earth $\lambda= 1 {\;}\textrm{light year}\approx 
1\cdot 10^{13}\,\textrm{km}$, and graviton energy is 
$\hbar \omega \approx 2\cdot 10^{-41}\textrm{J}=
1,3\cdot 10^{-22}\,\textrm{eV}$. \\

In quantum theory the power emitted by a planet is 
$P=n\hbar \omega$, where $n$ is the number of gravitons emitted 
within one second. However computing of how many gravitons are emitted 
by the planet within a second or in a shorter time interval makes no 
sense. The instant of graviton emission may be determined 
with accuracy $\Delta t$ not exceeding the wave period $T$, $\Delta t 
\geq T$, and for planets the periods are years (or at least days for extrasolar 
planets). Yet a physical meaning 
may be attributed to the energy $PT$ radiated away by the planet 
in the time interval equal to the wave period; this amount of energy 
is carried away by $N_g$ gravitons,  
\begin{displaymath}
N_g=\frac{PT}{\hbar \omega}.
\end{displaymath}
From (2) we get 
\begin{equation}
N_g = \frac{64\pi G^3}{5c^5\hbar}\frac{m^2M^2}{r^2}.
\end{equation}
In a table below we show for four planets: the emitted power 
$P$ (in watts), the common period $T$ of the waves and of the 
orbital motion and the number of gravitons $N_g$ emitted in the time 
interval $T$. If the radiation is viewed as a quantum process, 
the planets also cannot be treated as solid lumps with continuous 
mass density, they should be viewed as discrete systems of many 
nucleons. Let $N=m\slash m_p$ be the number of nucleons in a planet, 
where $m_p$ is proton mass. In the last row of the table the ratio  
$N_g\slash N$ of the two large numbers is given. \\

\begin{tabular}{|c|c|c|c|c|}\hline 
Planet  & Mercury   & Earth      & Jupiter      & Neptune \\
\hline
P [W]    &  70       & 200         & 5400     & $2,5\cdot 10^{-3}$ \\
T [years] &  0,24     &  1          &11,9       &165           \\
$N_g$    & $6\cdot 10^{48}$ & $3\cdot 10^{50}$ & $1\cdot 10^{54}$
 & $1\cdot 10^{50}$ \\
$N_g\slash N$  &  0,04     & 0,1       & 1     & $2\cdot 10^{-3}$ \\
\hline
\end{tabular}
\vspace{0.3cm}\\

The coincidence of the large numbers is evident and is particularly 
conspicuous in the case of Jupiter. The origin of the coincidence is 
unclear. It seems at present that the coincidence is rather 
accidental and there is no deeper reason for it to occur. 
Nevertheless it is amusing to find such a bizarre large number coincidence 
showing that the solar system reveals some strange regularities 
(another one is the Titius-Bode law). It is interesting to see if 
something similar occurs for planets orbiting other stars. The "Extrasolar 
Planets Catalog" \cite{Ext} contains 181 planets and out of them only 10 
planets have all relevant parameters measured: the star's and planet's mass, 
the inclination angle $i$, the eccentricity and the orbital period (or the 
semi-major axis). Among these eight planets move on almost circular 
orbits (the eccentricity is small). All these eight planets have masses of 
order of Jupiter mass, $1,9\cdot 10^{27}$kg, while their stars are of masses 
comparable to that of the Sun. This sample of planets is strongly 
observationally biased for obvious reasons: all the planets are very close 
to their stars, the radius of the largest orbit is only 0,20783 A.U., 
in all other cases it is smaller than 0,05 A.U. (less than $7\cdot 10^6$
km). As a 
consequence the ratio $N_g\slash N$ for the eight extrasolar planets varies 
from 100 for Gliese 876 b (period 60,9 days) to $8\cdot 10^4$ for 
OGLE-TR-56 b (period 1,21 days); in most cases it is of order $10^4$. The 
selection bias prevents one from inferring of whether the large number 
coincidence found in the Solar system is common among planetary systems or 
is exceptional. \\

\section{Is anything strange in the concept of graviton?}
From eq. (3) one derives 
\begin{equation}
\frac{N_g}{N} = \frac{64\pi G^3}{5c^5\hbar} m_p\frac{m M^2}{r^2}
\end{equation} 
and it is worth noticing that the ratio $N_g\slash N$ is a linear function 
of the moving mass $m$. If for solar planets it is of order 1, for 
light bodies $N_g\slash N << 1$. For a single hydrogen atom which 
lonely orbits around the Sun following the Earth's orbit the radiated 
power is $P \approx 2\cdot 10^{-101}\textrm{W}$ and
$N_g \approx 3\cdot 10^{-53}$, or this atom emits one graviton with 
$\lambda = 1{\;} \textrm{light year}$ in $3\cdot 10^{52}$ years. This 
means that according to quantum physics a hydrogen atom in this state of 
motion does not radiate at all. Yet if this atom is captured by the 
gravitational force of the Earth and falls into its atmosphere, the 
capability of the atom to radiate will grow many times though its 
macroscopic state of motion (which determines the emission power) has 
remained unchanged. In fact, when the hydrogen atom enters the atmosphere 
there appears a correlation with all the atoms of the planet 
since, according to eq. (2) $P\propto N_g \propto m^2$. 
The puzzle of the correlation lies in the fact that the mere effect of 
being bound to the Earth by its gravity makes the atom to radiate one 
graviton per ten years. A problem arises which---as far as we know--- 
up to now has not been clearly solved in the context of quantum gravity:
what is the physical mechanism of creating quantum gravity 
correlations between all atoms (or nucleons) in a macroscopic body 
which causes that the number of gravitons emitted by the body 
consisting of $N$ particles grows as $N^2$? \\

We emphasize that in classical linearized gravitational radiation theory 
the quadrupole formula and following from it eq. (2) for $P$ rise no doubts. 
Classically gravitational radiation is by definition a macroscopic effect 
and a whole planet acts as a single emitter, thus there is no problem of 
correlation between its atoms. The case of a planet orbiting around a star  
is slightly misleading because then all the formulae depend on 
the two masses and these enter eqs. (2) and (3) in different powers. It 
would be more clear to consider the case of two equal masses $m$ 
moving on a circular orbit around the center of mass with constant 
angular velocity under influence of a non-gravitational force or even 
one mass on a circular orbit. In the classical theory all particles of 
a given body give almost equal contributions to the amplitude of the wave 
and this implies that the emitted power is $P \propto m^2$. 
The problem appears only when one attempts 
to describe this macroscopic process as a combination of independent 
microscopic effects. \\

In the case of electromagnetic radiation (or quatum 
mechanical emission of any other elementary particles) the situation is 
different: each atom (molecule or nucleus) emits a photon 
independently of other atoms. As a result the number of photons 
emitted by a system of $N$ microscopic objects is proportional to $N$. 
It is exactly this property of electromagnetic interactions (as well as 
weak and strong interactions) that makes 
that quantum physics is physics of the microworld. Particles interact 
with other particles individually and not as collective systems, e.g. 
a neutrino coming from the space is captured by an individual nucleon 
in one atom in the Earth rather than by the entire planet. On the 
contrary, both emission and absorption of gravitons by a body is a 
kind of collective process arising due to the correlation between all 
particles of the body. This raises a question of whether 
in quantum gravity are there 
two--particle processes at all, such as elastic or inelastic scattering of 
(high energy) graviton on electron or proton or rather should one take into account 
a whole system of gravitationally bounded particles (i.e. it is the system 
that is subject to a quantum interaction with a single graviton)? \\

 To solve the problem we first compare the rotating 
quadrupole of two equal masses with the case of an electric dipole 
rotating with constant angular velocity. Let the dipole consist of $N$ 
charges $+e$ (grouped together) and $N$ charges $-e$ at a distance 
$l$. Each charge $+e$ interacts with each of $N$ negative charges 
and in this sense the dipole may be decomposed into $N^2$ pairs 
$+e-e$ and each such elementary dipole radiates independently of 
all others. One then expects that the total radiation power of the 
dipole is proportional to $N^2$. And in fact the power of classical 
electromagnetic dipole radiation is proportional to 
$((d^2/dt^2)\mathbf{d})^2$ where $\mathbf{d} =Ne\mathbf{l}$ is the 
electric dipole momentum, i.e. to $N^2$. 
Thus the classical pictures of electromagnetic and gravitational 
radiation from rotating systems are very similar. \\

The issue then is whether the quantum picture generates a real 
problem for gravitational radiation: appearance of quantum 
gravity correlations between nucleons in two bodies at large 
macroscopic distances. Are there similar effects for other 
interactions? In quantum field theory one usually investigates 
interactions between microscopic objects and these form our concepts 
about the quantum world. One can, however, also study interactions of 
classical systems with quantum fields, thus the problem 
is not specific to gravitation. Consider a 
macroscopic electric dipole, rotating or oscillating, coupled to 
quantum electromagnetic field. In QED a  
classical macroscopic current may be a source of the quantum field and 
in this case the number of generated photons is proportional to 
$N^2$, i.e. the classical and quantum computations do agree 
\cite{CT}. This 
means that the problem of macroscopic quantum correlations - if it 
exists at all - does appear already in QED. Yet in QED no one would 
admit that a radiating macroscopic electric dipole does generate a 
conceptual problem. \\
  May be the correct answer is that there is no physical problem at 
all and it is rather of psychological origin. QED is a well developed 
and fully reliable theory and if one does not get a satisfactory 
answer regarding the nature of quantum correlations at macroscopic 
distances in an electric dipole one does not interpret it as a 
defect of the theory. On the contrary, modern comprehending of QFT 
implies that search for a detailed physical picture of the process of 
quantum emission of radiation from macroscopic bodies is groundless.  
Such "physical pictures of quantum effects" do not correspond to 
anything real in the physical world and may be misleading. There 
is no deeper understanding of quantum processes beyond the 
outcomes of QFT calculations. There are no hidden parameters nor 
deeper insights in quantum world. Quantum theory of gravity is still 
in its very initial stage and we do not know which questions should be 
put forward and which should not. \\

\section{The issue of gravitons}
The conclusion that the classical quadrupole formula does not create 
paradoxes with the concept of graviton does not ensure that the concept 
itself is well grounded. The graviton is a fully quantum object, 
nevertheless its notion should be compatible with classical general 
relativity. \\
In QED the quantum electromagnetic field appears as a bundle of photons 
carrying energy and momentum of the field \cite{We1}. In the low energy 
limit the photon may be classically approximated by a plane 
electromagnetic wave. There is a clear one-to-one correspondence between 
photons and plane waves. In other terms one may precisely point to where 
in classical electrodynamics one makes the de Broglie conjecture 
providing transition to the quantum theory. By this we do not mean the 
standard generic procedure of quantizing a classical field theory. We do 
mean that in the classical electromagnetic field the plane waves are 
singled out and already in the classical theory one finds unambiguous 
hints that plane waves are related to quantum objects. In fact, consider 
the general explicitly relativistically invariant decomposition of the 
electromagnetic potential in plane waves with the wave vector $k^{\mu}$, 
\begin{equation}
A_{\mu}=\frac{1}{(2\pi)^{3/2}}\int_{-\infty}^{\infty} d^4k 
\left[a_{\mu}(k)e^{-ik_{\nu}x^{\nu}} + \overline a_{\mu}(k)
e^{ik_{\nu}x^{\nu}}\right]\delta(k_{\alpha}k^{\alpha})\theta (k^0),
\end{equation}
here $\overline a_{\mu}$ is the complex conjugate wave amplitude and 
$\theta (k^0)$ is the Heaviside step function (and the signature is 
$+---$). Performing integration over $k^0$ one gets the standard 
expression 
\begin{equation}
A_{\mu}=\frac{1}{2} \frac{1}{(2\pi)^{3/2}}\int_{-\infty}^{\infty} 
\frac{d^3k}{k^0} 
\left[a_{\mu}(\mathbf{k})e^{-ik_{\nu}x^{\nu}} + \overline a_{\mu}
(\mathbf{k})
e^{ik_{\nu}x^{\nu}}\right]
\end{equation}
with $k^0=|\mathbf{k}|>0$. Here the Lorentz invariant 3-dimensional 
integration measure is $d^3k/k^0$. The total energy and momentum of 
the field is 
\begin{displaymath}
P^{\mu}=\frac{1}{c} \int_{\mathbf{R}^3} T^{0\mu}\, d^3x,
\end{displaymath}
where $T^{\mu\nu}$ is the symmetric (gauge invariant) 
energy-momentum tensor (we use conventions of ref. \cite{LL}). 
Inserting (6) one replaces the integral over 
the whole physical space with an integral over the momentum 
3-space\footnote{It is worth noticing that one cannot obtain an 
analogous Lorentz covariant formula using the spectral decomposition 
of the field strength instead of the potential.}
\cite{SSS}
\begin{equation}
P^{\mu}=\frac{-1}{8\pi c} \int_{-\infty}^{\infty} 
\frac{d^3k}{k^0}\, 
a_{\nu}(\mathbf{k})\overline a^{\nu}(\mathbf{k})\,k^{\mu}.
\end{equation}
The Lorentz gauge condition requires $a_{\nu}k^{\nu}=0$ what implies, 
since $k^{\mu}$ is null, that both the real and imaginary parts of the 
amplitude vector are spacelike vectors (if they are null they are 
gradients), then $a_{\nu}\overline a^{\nu}
<0$. Therefore the quantity 
\begin{displaymath}
\frac{-1}{8\pi c} \frac{d^3k}{k^0} 
a_{\nu}(\mathbf{k})\overline a^{\nu}(\mathbf{k})
\end{displaymath}
is a Lorentz scalar which is positive and vanishes only for $A_{\mu}=0$ 
and has dimension $\textrm{M}\textrm{L}^2
\textrm{T}^{-1}$, i.e. the dimension of Planck constant. It is 
here that one makes the Planck--Einstein--de Broglie conjecture: 
the quantity is equal 
to $n(\omega) d\omega\, \hbar$, where $\hbar$ is a new universal 
dimensional constant signalling transition to quantum effects and 
$n(\omega) d\omega$ is a number of quantum "particles" (photons) 
determined (and denoted) by the wave vector $\mathbf{k}$ where the 
values of $\mathbf{k}$ belong to the interval $[k, k+dk]$ with 
$k=k^0=|\mathbf{k}|=\omega/c$. Then 
\begin{equation}
P^{\mu}= \int_{0}^{\infty} n(\omega) d\omega\, \hbar k^{\mu}
\end{equation}
and the total energy and momentum of the field is interpreted as a 
sum over all quantum particles, each carrying four-momentum equal 
$\hbar k^{\mu}$. In this sense, apart from and independently of the 
quantization formalism, there is a direct and physically clear 
relationship between classical and quantum electromagnetic fields. \\

In linearized general relativity the relationship between the classical 
and quantum fields may be derived in an analogous way though there are 
some troubles. There are two approaches to the problem. The first 
approach is based on the use of the field equations alone. A metric 
perturbation $g_{\mu\nu}=\eta_{\mu\nu}+h_{\mu\nu}$ around Minkowski 
space gives rise to the linearized Riemann tensor which is gauge 
invariant and applying the harmonic gauge condition 
$h^{\mu\nu}{}_{,\nu}=0$ and $h\equiv \eta^{\mu\nu}h_{\mu\nu}=0$ 
the field equations are reduced to 
$\Box h_{\mu\nu}\equiv\partial^\alpha\partial_\alpha
h_{\mu\nu}=0$. Analogously to the electromagnetic case a general 
solution in this gauge is 
\begin{equation}
h_{\mu\nu}=\frac{1}{2} \frac{1}{(2\pi)^{3/2}}\int_{-\infty}^{\infty} 
\frac{d^3k}{k^0} 
\left[a_{\mu\nu}(\mathbf{k})e^{-ik_{\alpha}x^{\alpha}} + \overline 
a_{\mu\nu}(\mathbf{k})e^{ik_{\alpha}x^{\alpha}}\right]
\end{equation}
with $k^{\mu}k_{\mu}=0$ and the wave amplitude (the polarization 
pseudotensor) is restricted by $a_{\mu\nu}k^{\nu}=0$ and 
$\eta^{\mu\nu}a_{\mu\nu}=0$. One then introduces a Lorentz covariant 
energy--momentum pseudotensor $t_{\mu\nu}$ which is the second--order 
part in $h_{\mu\nu}$ of the Ricci tensor, see e.g.\/ \cite{We2}. For a 
single monochromatic wave the pseudotensor averaged over a spacetime 
region of size much larger than $|\mathbf{k}|^{-1}$ is 
\begin{equation}
\langle t_{\mu\nu}\rangle =\frac{c^4}{16\pi G} k_{\mu}k_{\nu}
a_{\alpha\beta}\overline a^{\alpha\beta}.
\end{equation}
The pseudoscalar $a_{\alpha\beta}\overline a^{\alpha\beta}$ is 
nonnegative, in fact, by a Lorentz transformation one may get 
$k^{\mu}= k^0 (1, -1, 0, 0)$ and the harmonic gauge implies 
$a_{\mu 1}=a_{\mu 0}$ and $a_{33}=-a_{22}$. The polarization 
pseudotensor has then 5 independent components and under the 
remaining gauge transformations, $a_{\mu\nu} \to a_{\mu\nu} + 
\epsilon_{\mu}k_{\nu} + \epsilon_{\nu}k_{\mu}$, where 
$\epsilon_{\mu}(\mathbf{k})$ is a Lorentz covariant complex 
vector subject to $\epsilon_{\mu}k^{\mu}=0$, three of them are 
changed and may be set equal to zero and only the other two, 
$a_{22}$ and $a_{23}$ remain gauge invariant. The helicity $\pm 2$ 
plane wave is described by $a_{22}$ and $a_{23}$. One gets 
\begin{displaymath}
a_{\alpha\beta}\overline a^{\alpha\beta} = 2|a_{22}|^2 + 
2|a_{23}|^2>0
\end{displaymath}
and this fact allows one to assign the number $n$ of gravitons, each 
having the four--momentum $\hbar\, k^{\mu}$, to unit volume of the 
plane wave \cite{We2},
\begin{equation}
n =\frac{c^2}{8\pi G \hbar} \omega (|a_{22}|^2 + |a_{23}|^2).
\end{equation}
The other approach seems more reliable since one formulates the 
linearized general relativity as a Lagrangian theory for 
a massless spin--2 field in flat spacetime. To this end one  
generates a Lagrangian for the field by taking the second variation of 
Einstein--Hilbert Lagrangian with respect to the metric perturbation 
$h_{\mu\nu}$ around the flat background 
\cite{AD}. The resulting Lagrangian appeared first in the textbook 
\cite{W} and will be referred to as Wentzel Lagrangian,
\begin{equation}
L_W =\frac{1}{4}\left(-h^{\mu\nu;\alpha}h_{\mu\nu;\alpha}
+2h^{\mu\nu;\alpha}h_{\alpha\mu;\nu}
-2h^{\mu\nu}{}_{;\nu}h_{;\mu}+h^{;\mu}h_{;\mu}\right)
\end{equation} 
with $h\equiv g^{\mu\nu}h_{\mu\nu}$ and $g_{\mu\nu}$ is the flat spacetime 
metric in arbitrary coordinates. The Lagrangian and the field equations 
are gauge invariant and again 
the harmonic gauge $h^{\mu\nu}{}_{;\nu}=0=h$ is most convenient (the 
covariant derivatives are introduced for later use). 
As is well known, the theory is defective for there is no gauge 
invariant energy--momentum tensor for $h_{\mu\nu}$: both the canonical 
\cite{Tr} and the variational (metric) energy--momentum tensors 
\cite{AD, DMC} are gauge dependent and cannot be improved. This 
is a particular case of a generic situation where the variational 
energy--momentum tensor (hereafter denoted as the stress tensor) does not 
inherit the symmetries of the underlying Lagrangian \cite{MS1}. As a 
consequence it is impossible to attach a physical meaning to the 
local distribution of energy already for a linear spin--2 field. Only in 
special cases, e.g.\/ for plane waves, one can use a particular gauge  
and then the canonical energy--momentum tensor may be invariant under 
the remaining gauge transformations \cite{Tr}. This notion of local 
energy has, however, a limited meaning. \\

In general relativity the (matter) stress tensor acts as the source 
of gravity and one expects that it plays a distinguished role also for 
the linearized gravity itself, though it is not gauge invariant. 
Wentzel Lagrangian is expressed in terms of covariant 
derivatives and formally assuming that the metric $g_{\mu\nu}$ in (12) 
is arbitrary, one may use the standard definition to calculate the 
tensor. It is very complicated \cite{MS1} and we apply the 
harmonic gauge condition to simplify it, then it reads 
\begin{eqnarray}
T_{\mu\nu}^{W}(h,\eta)&=&
-h_{\mu\nu;\alpha\beta}h^{\alpha\beta}
-2h_{\alpha\beta;(\mu}h_{\nu)}{}^{\alpha;\beta}
+\frac{1}{2}h_{\alpha\beta;\mu}h^{\alpha\beta}{}_{;\nu}
+2h_{\mu}{}^{\alpha;\beta}h_{\nu(\alpha;\beta)}
 \nonumber\\ &&
+\frac{1}{4}g_{\mu\nu}\left(-h^{\alpha\beta;\sigma}h_{\alpha\beta;
\sigma}
+2h^{\alpha\beta;\sigma}h_{\sigma\alpha;\beta}
\right).
\end{eqnarray}
This expression holds only in
flat spacetime since in deriving it one assumes that the covariant
derivatives commute. In Cartesian coordinates the general solution for 
$h_{\mu\nu}$ is given by eq. (9) and the total four--momentum for the 
field is 
\begin{equation}
P^{\mu}= \frac{1}{c} \int_{\mathbf{R}^3} T^{W0\mu}\, d^3x =
\frac{1}{4 c} \int_{-\infty}^{\infty} 
\frac{d^3k}{k^0}\, 
a_{\alpha\beta}(\mathbf{k})\overline a^{\alpha\beta}(\mathbf{k})\,
k^{\mu}.
\end{equation}
Clearly this formula agrees with (10) and (11) for a suitably chosen 
normalization factor in $T^{W}_{\mu\nu}$. These expressions are 
invariant under the remaining gauge transformations 
$a'_{\mu\nu} = a_{\mu\nu} + 
\epsilon_{\mu}k_{\nu} + \epsilon_{\nu}k_{\mu}$. Furthermore, Deser and 
McCarthy \cite{DMC} have proved a "folk theorem" to the effect that 
under an arbitrary gauge transformation of a gauge invariant quadratic 
Lagrangian (i.e.\/ $L_W$ or any other equivalent to it in Minkowski space) 
the stress tensor computed "on shell" (the field equations hold) is 
varied only by superpotential terms. A superpotential term, by its 
construction, gives no contribution to the Poincar\'e generators. In 
other terms the Poincar\'e generators are gauge invariant. This means 
that the total $P^{\mu}$ evaluated in the harmonic gauge using eq. (14) 
is equal to the four--momentum evaluated in any other gauge or without 
any gauge (beyond the harmonic gauge eqs. (9)--(11) and (13)--(14) are 
not valid). The gauge invariance of the total $P^{\mu}$ together with
the gauge dependence of the stress tensor implies that while $P^{\mu}$ 
is numerically the same when computed in any gauge, the integrand in 
each case is different from that in the second integral in (14). When 
$P^{\mu}$ is computed in classical electrodynamics in various gauges, 
in each case the integrand has a form different from that in eq. (7) 
but its value for a given $\mathbf{k}$ is always the same. Yet in 
the linearized gravity the integrand has both different forms and 
different values for various gauges. This is why the standard 
interpretation \cite{We2} of (11) as the number of gravitons with the 
momentum $\hbar\,k^{\mu}$ raises doubts. \\

The exact nonlinear general relativity is different in this aspect from 
any linear field theory. Due to the principle of equivalence the notion 
of local distribution of energy is meaningless (this is a stronger cause 
than the gauge non--invariance in the linearized gravity) and one should 
search for the relationship between the quantum and the classical theory 
in a different way. It is commonly accepted that exact purely radiative  
fields are described by the plane--fronted gravitational waves with 
parallel rays (\emph{pp} waves) \cite{JEK}. Comparing the Weyl tensor 
for \emph{pp} waves with the Maxwell field strength for a plane 
electromagnetic wave one finds similarities permitting analogous 
physical interpretation \cite{JEK, SKMHH}. 
A subclass of \emph{pp} waves, 
the plane wave manifolds, are geodesically complete and for this reason 
are regarded as classical analogues of gravitons \cite{JEK, GR}. 
However energy considerations indicate that this interpretation is not 
well grounded. Although the \emph{pp} waves are exact solutions of the 
nonlinear vacuum Einstein field equations, they also constitute their own 
linear approximation (sometimes this feature is used as an argument 
supporting that these spacetimes correspond to gravitons). As a 
consequence the total energy assigned to these solutions is always zero. 
In fact, in the ADM approach to gravitational energy of a solution one 
introduces a 
pseudotensor defined as the quadratic and higher order terms part in the 
decomposition of Ricci tensor in the metric perturbations $h_{\mu\nu}$. 
A physical meaning has only the total ADM energy being a surface integral 
of the pseudotensor over a sphere at spatial infinity (assuming that the 
spacetime is asymptotically flat there). For the 
\emph{pp} waves the full Ricci tensor is equal to its linear 
approximation, hence the pseudotensor and the total ADM energy are zero 
(regardless of that the plane waves are not asymptotically flat). One 
cannot view plane gravitational waves as a classical low energy 
approximation to a swarm of quantum particles carrying energy. \\

In ordinary quantum mechanics one associates with any microscopic object 
having rest mass and energy $E$ (elementary particle, nucleus, atom) a 
quantum wave of frequency $\omega$ and assumes $E=\hbar \omega$. 
Contrary to the case of electromagnetic field, eqs. (7) and (8), one 
cannot recognize this frequency on classical grounds. In fact, if a 
microscopic system has a classically defined length scale $\lambda$ it 
may be used to formally define a frequency $\omega_{\mathrm{cl}}=2\pi c/
\lambda$; this frequency, however, is unrelated to the true quantum 
frequency $E/\hbar$. For instance, the classical electron radius 
$\frac{e^2}{mc^2}$ is some hundred times smaller than the quantum 
electron Compton wavelength $\hbar/mc$. One expects that the same holds 
for these gravitational fields which are analogous in a sense to 
classical particles. A black hole is essentially an elementary object 
which cannot be decomposed into its constituent parts and is as 
fundamental as the electron. The static black hole of mass $M$ defines 
its length scale, the horizon radius $r_g =\frac{2GM}{c^2}$, and the 
classical frequency $\omega_{\textrm{cl}}=\frac{\pi c^3}{GM}$ is 
assigned to it. Since the length scale is not independent from the 
black hole energy $E$ a bizarre relation arises, 
\begin{displaymath}
E = \frac{\pi c^5}{G\omega_{\mathrm{cl}}}; 
\end{displaymath}
clearly $\omega_{\mathrm{cl}}$ has nothing to do with a hypothetical 
quantum frequency which might be associated with the black hole in 
quantum gravity. \\

In standard quantum field theory the particle--wave duality establishes 
a universal correspondence between energy (mass) and frequency. This 
correspondence 
is inconsistent with classical general relativity. In Einstein's theory 
(and in other metric theories of gravity) a good notion of energy is 
provided only by the total ADM energy. This notion of energy has a unique 
status in general relativity. In no other theories of physics is energy 
effectively a charge and the same holds for momentum. In classical 
electrodynamics, where all the relevant calculations can explicitly be 
done, the electromagnetic field of a system of charges can be 
decomposed at the spatial infinity into the radiative field which gives no 
contribution to the total charge as being transversal and the "Coulomb 
field" which falls off as $r^{-2}$. The latter can be spectrally 
decomposed into longitudinal waves with zero frequency \cite{LL}. Thus 
the electric charge is universally associated with classical zero 
frequency waves. We emphasize that in general relativity 
the Poincar\'e generators being charges are calculated as 
integrals over a two--sphere at the spatial infinity. This means that the 
energy and momentum of an asymptotically flat spacetime are 
zero frequency characteristics of this spacetime. In consequence this 
implies that any general relationship between energy and frequency is 
\emph{a priori\/} impossible and in particular the Einstein formula 
$E=\hbar \omega$ cannot hold. \\

\section{Conclusions}
The standard notion of the graviton as a quantum of the radiation 
gravitational field, which is akin to the photon, may be justified (with 
some reservation) only in the framework of the linearized gravity (which 
itself is a defective theory). In this theory it gives rise to an 
amusing large number coincidence between the number of gravitons 
radiated away by solar planets in a time interval equal to the wave 
period and the number of nucleons in these planets. In full (nonlinear) 
general relativity the (ADM) energy is, for fundamental reasons, unrelated 
to wave phenomena, in particular it is disconnected from the wave frequency. 
This statement does not mean that gravitons do not exist. If 
gravitons as quanta of a quantum gravitational field do exist, their 
properties are different from those of photons in QED. The nature of 
gravitons may be determined only in the framework of full quantum theory 
of gravity and without knowing it one can say nothing about them. The case 
of electromagnetism is misleading in this aspect: either classical 
general relativity bears no traces of quantum effects at all or the traces 
of gravitons are different from those of photons and at present are 
unrecognizable in the theory. In particular one cannot expect that in 
the low energy classical approximation the gravitons may be interpreted 
as the \emph{pp} waves.\\
Here it has been assumed that the gravitational field carries a 
fundamental interaction and as such is of quantum nature. If gravity is 
an "emergent" phenomenon, i.e.\/ if it arises as a kind of averaging of 
other elementary interactions \cite{BLV}, then quantization of it makes no 
sense.\\

The authors are grateful to Henryk Arod\'z, Leonid Grishchuk and Jakub 
Zakrzewski for helpful comments and discussions.

\end{document}